\documentclass[a4paper,11pt]{article}
\usepackage{jinstpub} 
\usepackage{lineno}
\usepackage{siunitx}
\usepackage{orcidlink}
\usepackage{mathtools}
\usepackage{wrapfig}

\title{\boldmath The vertexing challenge at FCC-ee}

\collaboration[c]{\normalsize on behalf of FCC}

\author[a,1]{Armin Ilg\note{Corresponding author.}\orcidlink{0000-0001-9488-8095}}
\author[a]{Anna Macchiolo\orcidlink{0000-0003-0199-6957}}
\author[b]{Fabrizio Palla\orcidlink{0000-0002-6361-438X}}
\affiliation[a]{Physik Institut, University of Zürich\\
Winterthurerstrasse 190, 8057 Zürich, Switzerland}
\affiliation[b]{INFN Pisa\\
Largo Bruno Pontecorvo 3, 56127 Pisa, Italy}

\emailAdd{armin.ilg@cern.ch}

\abstract{Following in the footsteps of the LHC, the Future Circular Collider (FCC) plans to be the next multi-generational international collider project. In the first stage, FCC-ee will collide intense beams of electrons and positrons at centre of mass energies between 87 and \SI{365}{GeV}, making it an electroweak, flavour, Higgs and top factory. The unprecedented statistical precision requires FCC-ee experiments to limit their systematic uncertainties to the very minimum.

The precise reconstruction of the interaction vertices is central to most measurements at FCC-ee, such as rare flavour physics processes and the measurement of Higgs and $Z$ decays to bottom and charm quarks and taus.
This contribution will discuss the requirements of FCC-ee vertex detectors, from the necessary impact parameter resolution via the challenging collision environment at the Z pole to the tight requirement on the material budget, which should be kept below $\approx 0.3\%$ of a radiation length per detection layer. Next, the proposed vertex detector designs for FCC-ee are shortly presented, and an outlook is given on novel detector designs and features.

The requirements for the vertexing performance translate into requirements for the sensors used for the vertex detector. As discussed in this contribution, they need to feature a spatial resolution of about \SI{3}{\micro\meter} and provide timing information of $\mathcal{O}(\SI{}{\nano\second} \text{-} \SI{}{\micro\second})$ while keeping power consumption minimal to allow for air-cooling of the detector – minimising the detector material budget.
The only type of sensor capable of aiming to fulfil such requirements are CMOS Monolithic Active Pixel Sensors (MAPS), which combine signal generation, amplification and readout into a single silicon die. Therefore, the rest of this contribution will present an overview of existing and planned MAPS technologies and prototypes aiming to fulfil the stringent FCC-ee vertex detector requirements.
}

\keywords{Particle tracking detectors (Solid-state detectors), large detector systems for particle and astroparticle physics}

\arxivnumber{2502.04071} 

\notoc

\begin{document}
\maketitle
\flushbottom

\section{The Future Circular Collider}

The FCC is a \SI{90.7}{km} circumference collider infrastructure proposed to be built at CERN to continue the legacy of the LHC. The first incarnation, the FCC-ee \cite{FCC_FS_PED,FCC_FS_Acc}, is a high-intensity $e^+ e^-$ collider scheduled to start shortly after the HL-LHC, around 2045. Collisions at centre-of-mass energies from $\sim 87$ up to \SI{365}{GeV} enable the detailed study of the Higgs boson, electroweak physics, and top physics. Through an expected dataset of about $10^{12}$ $b\bar{b}$ and $c\bar{c}$ pairs as well as $\SI{2e11}{}$ pairs of $\tau$ leptons, the four-year long run at the Z pole will also make FCC-ee an excellent flavour physics facility. It will feature four interaction points (IPs) to host experiments. The civil engineering infrastructure is compatible with FCC-hh, a hadron-hadron collider that could later deliver collisions at energies of $\sim \SI{100}{TeV}$. 

The challenge faced by the FCC-ee is matching the tiny statistical uncertainties, mainly resulting from extensive statistics of the Z pole run, with equally small theoretical and experimental uncertainties. This is necessary to crack open the mysteries of particle physics by measuring and mapping the properties of the Standard Model (SM) of particle physics more precisely and finding tiny deviations from predictions.

\section{Vertex reconstruction at FCC-ee}

The efficient and precise reconstruction of the locations of the primary, secondary and tertiary decay vertices is crucial for most of the FCC-ee physics programme. Examples are the measurement of the tau lifetime, the measurement of the Higgs coupling to the bottom and charm quarks, the flavour physics measurements, and the impact on flavour tagging\footnote{The identification of the original type of quark producing a collimated stream of particles, a jet.} performance. All FCC-ee detector concepts thus foresee dedicated vertex detectors (VXDs) at their centres.

The impact parameter resolution is often used as a metric to gauge the vertexing capabilities. The currently assumed required resolution on the transverse impact parameter
is $\SI{3}{\micro\meter} \oplus \SI{15}{\micro\meter\giga\electronvolt}/(p \sin^{3/2} \theta)$. The asymptotic term is mainly governed by the sensor's spatial resolution and the distance to the IP. A small outer beam pipe radius of \SI{11.7}{mm} at FCC-ee allows to place the first VXD layer at $r \approx \SI{13.7}{mm}$, greatly helping to achieve the \SI{3}{\micro\meter} resolution. The second term describes the effect of multiple Coulomb scattering of the particles in the detector and beam pipe, deteriorating the resolution. 
A small number of radiation lengths (low \textit{material budget}) in the beam pipe and VXD is vital for reconstructing complex flavour physics decay chains that feature particles of comparatively low momentum at the Z pole. The foreseen AlBeMet162\footnote{Alloy of Aluminium (38\%)
and Beryllium (62\%)} beam pipe features only $\SI{0.7}{\percent}\ X_0$ \cite{MDI_2024} and the VXD designs target about $\SI{0.3}{\percent}\ X_0$ per layer \cite{Ilg2024}. The sensor thickness should be minimised, and lightweight support structures should be deployed to achieve this. The sensors are air-cooled, which requires a power consumption of $\lesssim \SI{50}{\milli\watt\per\centi\meter\squared}$ of the sensors. Achieving the aspired impact parameter resolution also depends on operational aspects such as the precise knowledge of the detector position and the alignment with the rest of the machine and detector. 

\begin{table}[htbp]
    \caption{Global requirements of FCC-ee vertex detectors.}
    \label{tab:global_req}
    \begin{tabular}{@{}p{0.37\textwidth}|p{0.62\textwidth}@{}}
        \textbf{Physics challenges} & \textbf{Requirement} \\ \hline
        Coverage up to $\left| \cos\theta \right| = 0.99$ & Long barrel, forward disks \\ \hline
        High reconstruction efficiency & Hermeticity, small peripheries, $>99\%$ hit eff., many layers \\ \hline
        Asymptotic resolution of $a \approx \SI{3}{\micro\meter}$ & \SI{3}{\micro\meter} single-hit resolution, first layer close to interaction point \\	\hline
        & Light beam pipe \\
        Multiple scattering: $b \approx \SI{15}{\micro\meter\GeV}$ & $\leq \SI{0.3}{\percent}\ X_0$/layer $\rightarrow$ thin sensors, light support \\
        & Air-cooling $\rightarrow$ Power consumption $\lesssim \SI{50}{\milli\watt\per\centi\meter\squared}$ \\
    \end{tabular}
\end{table}

In addition to this resolution requirement, the vertex reconstruction efficiency must be high, and no tracks shall escape the VXD unnoticed. A large forward coverage is desired (up to $\left|\cos \theta \right| = 0.99$). A complex integration of the VXD with the \textit{machine-detector interface} (MDI) region is necessary since the closest machine elements are inside the detectors \cite{MDI_2024}.

The global VXD performance requirements are summarised in \autoref{tab:global_req}.

\section{Collision environment at the Z pole}

The main challenge of FCC-ee VXDs comes from the operation at the Z pole. The instantaneous luminosity is foreseen to reach \SI{140e34}{\per\centi\meter\square\per\second} which is achieved through a bunch crossing rate of \SI{50}{\mega\hertz}, leading to a bunch spacing of \SI{20}{\nano\second}. This necessitates keeping the on-detector electronics powered on continuously, which makes the power consumption requirement for air-cooling difficult to fulfil. The integration time should be $\lesssim \SI{1}{\micro\second}$ to suppress multiple Z bosons during one integration time. The benefits of time resolutions below this need to be evaluated in the coming years. 

The large rate of $\SI{60}{kHz}$ of Z bosons, together with a significant number of $\gamma \gamma \rightarrow \text{hadrons}$ and Bhabha pairs, requires the detector to readout the events at a frequency greater than $\SI{100}{kHz}$. If no trigger is to be used, the VXD will be read out at the collision rate of \SI{50}{\mega\hertz} -- a challenge given the tight requirement on power consumption. The beam backgrounds at the Z pole are dominated by incoherent pair production and synchrotron radiation. Incoherent pairs are currently estimated to drive the particle hit rate in the VXD (\SI{200}{\mega\hertz\per\centi\meter\squared} as to author's evaluation\footnote{IDEA vertex detector in 2023 FCC-ee lattice} and \SI{370}{\mega\hertz\per\centi\meter\squared} in Reference~\cite{tuzat}). The impact of synchrotron radiation and other backgrounds is being studied and could further increase this hit rate requirement. These beam backgrounds also determine the radiation environment with which the VXD must cope. Radiation levels of few $\SI{e13}{}\ \SI{1}{MeV}\ n_\text{eq}\SI{}{\per\centi\meter\squared}$ and few tens of $\SI{}{\kilo\gray}$ per year are currently expected from the dominant radiative Bhabha and incoherent pair production processes \cite{FCC_FS_PED}.

\autoref{tab:environment} summarises the discussed requirements at the Z pole run.

\begin{table}[htbp]
    \caption{Requirements of FCC-ee vertex detectors at the Z pole.}
    \label{tab:environment}
    \begin{tabular}{@{}p{0.355\textwidth}|p{0.63\textwidth}@{}}
        \textbf{Collision environment challenges} & \textbf{Requirement} \\ \hline
        High luminosity & Readout between \SI{50}{\mega\hertz} (no trigger) and $\gtrsim \SI{100}{\kilo\hertz}$ (trigger) \\ \hline
        Avoid pile-up of Z bosons & Integration time $\lesssim \SI{1}{\micro\second}$ \\ \hline
        Beam backgrounds & Hit rate capability up to $\mathcal{O}(\SI{200}{\mega\hertz\per\centi\meter\squared})$ \\ \hline
        Radiation environment & Few $\SI{e13}{}\ \SI{1}{MeV}\ n_\text{eq}\SI{}{\per\centi\meter\squared}$ and few tens of $\SI{}{\kilo\gray}$ per year
    \end{tabular}
\end{table}

\section{Proposed vertex detector layouts and performance}

Four detector concepts are currently proposed for FCC-ee: IDEA, CLD, ALLEGRO, and ILD@FCC-ee. Two original VXD layouts exist from IDEA and CLD and they are currently also used in ALLEGRO and ILD@FCC-ee. All VXDs foresee the usage of Monolithic Active Pixel Sensors (MAPS) since they can offer a small pixel pitch, low material budget, and are cost-effective. 

\begin{figure}[htbp]
    \centering
    \begin{minipage}[b]{0.58\textwidth}
        \centering
        \includegraphics[width=\textwidth]{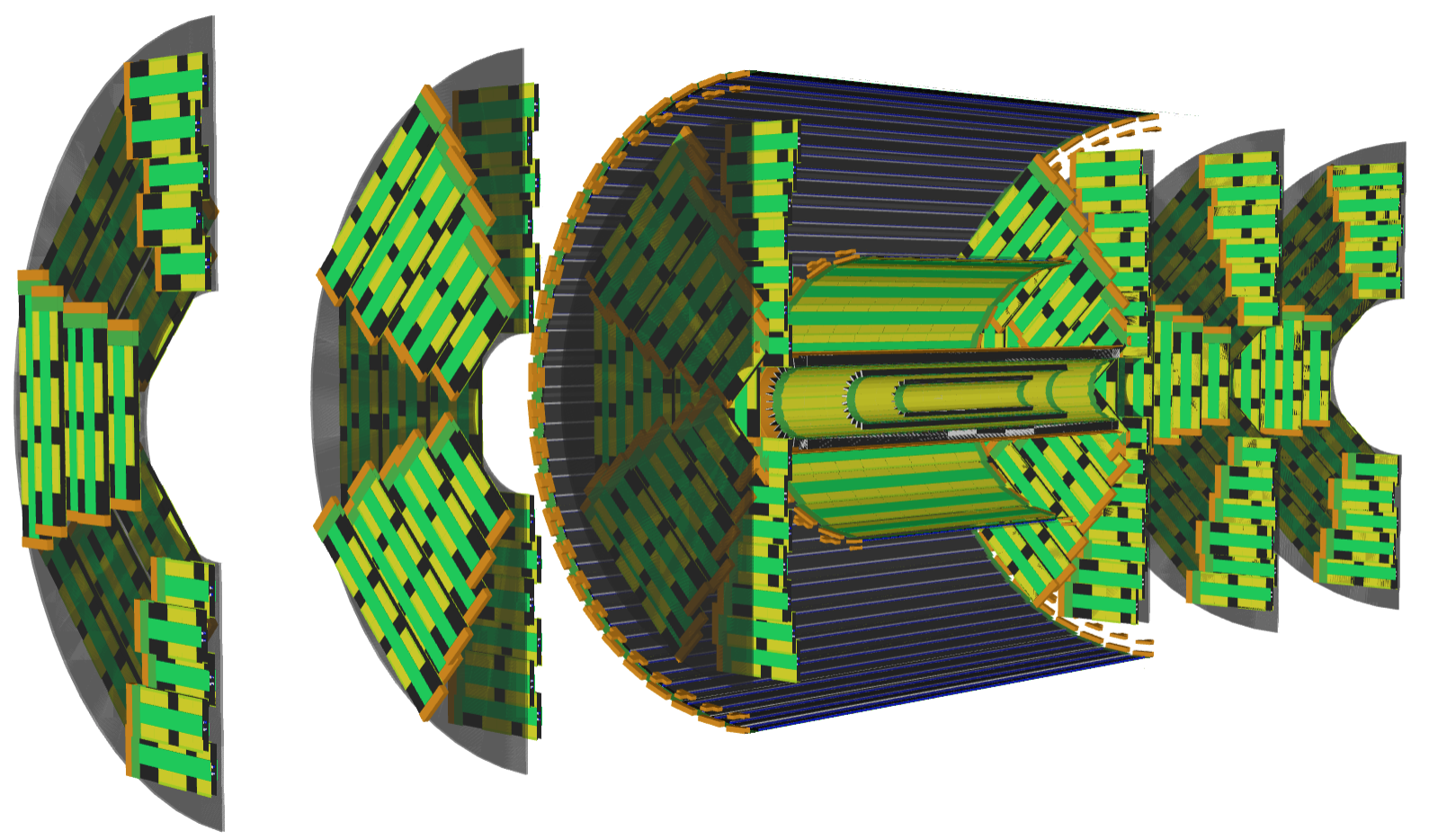}
        \caption{IDEA vertex detector full simulation model \cite{Ilg2024}.}
        \label{fig:IDEA_vertex}
    \end{minipage}\hfill
    \begin{minipage}[b]{0.41\textwidth}
        \centering
        \includegraphics[width=\textwidth,trim=0.8cm 0.8cm 0.8cm 0.8cm,clip]{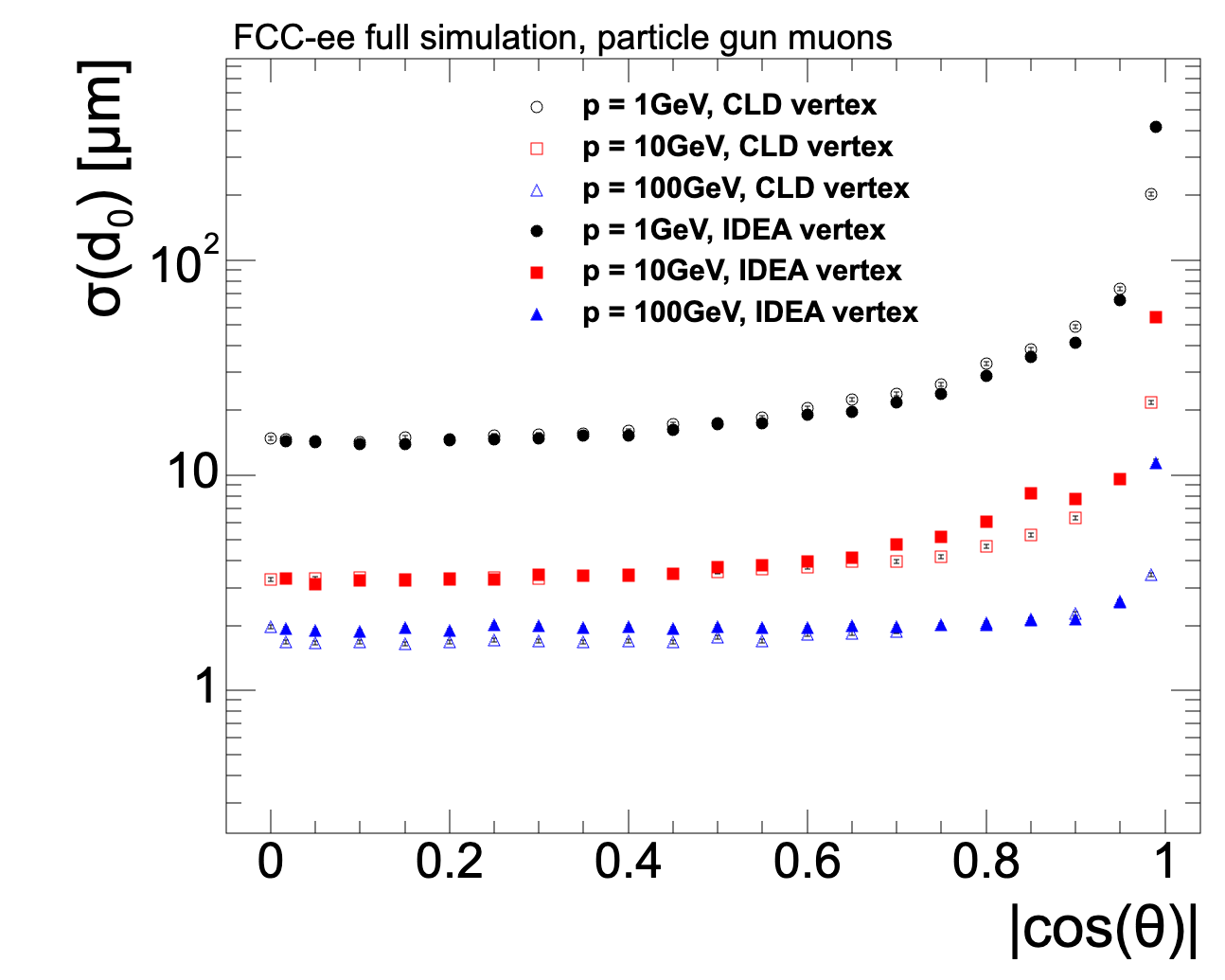}
        \caption{Transverse impact parameter resolution in CLD and IDEA VXD \cite{Ilg2024}.}
        \label{fig:impact_parameter}
    \end{minipage}
\end{figure}

CLD foresees three double-sided barrel layers and disks per side, with the first layer being at a radius of only \SI{12.5}{mm} \cite{CLD}. For the IDEA VXD, a first engineering design exists and is integrated into the MDI region of FCC-ee. It entails three single-hit inner vertex barrel layers with staves of ARCADIA \cite{Neubser2023} modules at a radius between 13.7 and  \SI{34}{\milli\meter}. The outer vertex consists of large-area MAPS forming an intermediate and outer barrel layer up to \SI{315}{mm} in radius and three disks per side up to \SI{960}{mm} in $z$. Refer to Reference~\cite{MDI_2024} for more details on the engineering. \autoref{fig:IDEA_vertex} shows the layout of the IDEA VXD in full simulation. The achievable transverse impact parameter resolution for both the CLD and IDEA VXDs is displayed in \autoref{fig:impact_parameter}. They both offer an asymptotic resolution of $< \SI{2}{\micro\meter}$ at $\cos\theta = 0$ for high-momenta particles and a multiple Coulomb scattering term of $\approx \SI{15}{\micro\meter\giga\electronvolt}$.

\section{Going beyond current vertexing capabilities}

There are many merits to going beyond the currently proposed and required performance of VXDs at FCC-ee. \autoref{tab:advanced} lists some potential advancements for FCC-ee VXD discussed in the following.

A factor of $\approx 2$ improvement in the impact parameter resolution\footnote{Compared to the requirement of $\SI{3}{\micro\meter} \oplus \SI{15}{\micro\meter\giga\electronvolt}/(p \sin^{3/2} \theta)$ on $\sigma_{d_0}$, for $\mathcal{O}(\SIrange{0.5}{5}{\giga\electronvolt/c})$ tracks} could, for example, open the possibility to measure the process $B^0 \rightarrow K^{*0} \tau^+ \tau^-$ down to the SM branching ratio expectation of $\mathcal{O}(10^{-7})$ \cite{Miralles2024}. This would shed light on the potential enhancement of this process, as expected in many beyond the SM theories \cite{Kamenik2017}. To achieve this, the material budget must be reduced.

\begin{wrapfigure}[5]{r}{0.64\textwidth}
    \centering    
    \vspace{-0.8\baselineskip}
    \includegraphics[width=\linewidth,trim=0.5cm 1.8cm 0.5cm 1.2cm,clip]{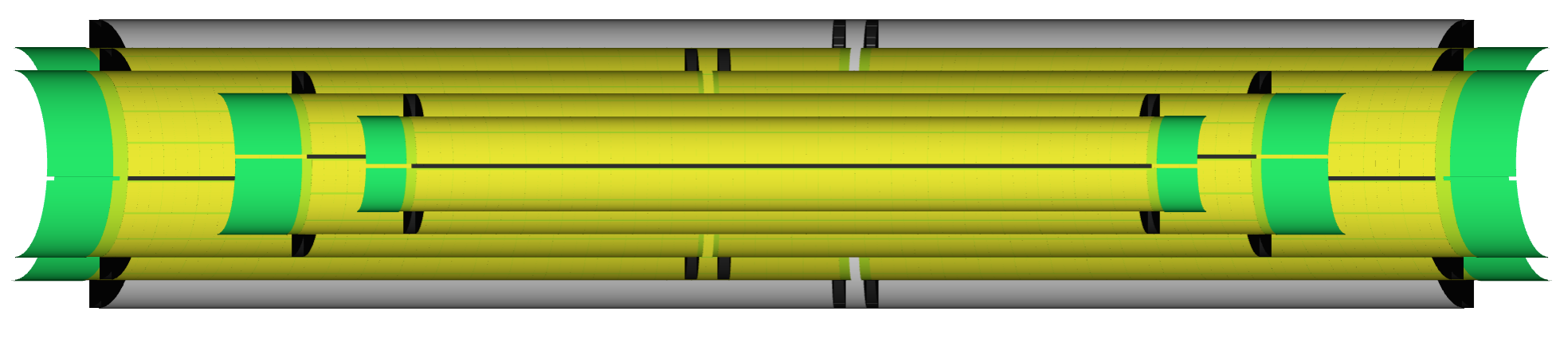}
    \caption{Cross-section of the ultra-light FCC-ee VXD concept.}
    \label{fig:ultra-light}
\end{wrapfigure}

One option is to follow the developments of ALICE ITS3 \cite{its3_tdr} to use stitched, wafer-scale MAPS that are thinned to about \SI{50}{\micro\meter} and bent to form half-cylinders. In this way, almost all support structures and cabling can be dismissed or moved to the edge of the detector acceptance. 

\begin{wrapfigure}[14]{l}{0.38\textwidth}
    \vspace{-0.5\baselineskip}
    \centering    
    \includegraphics[width=\linewidth,trim=0.1cm 0.3cm 0.6cm 0.7cm,clip]{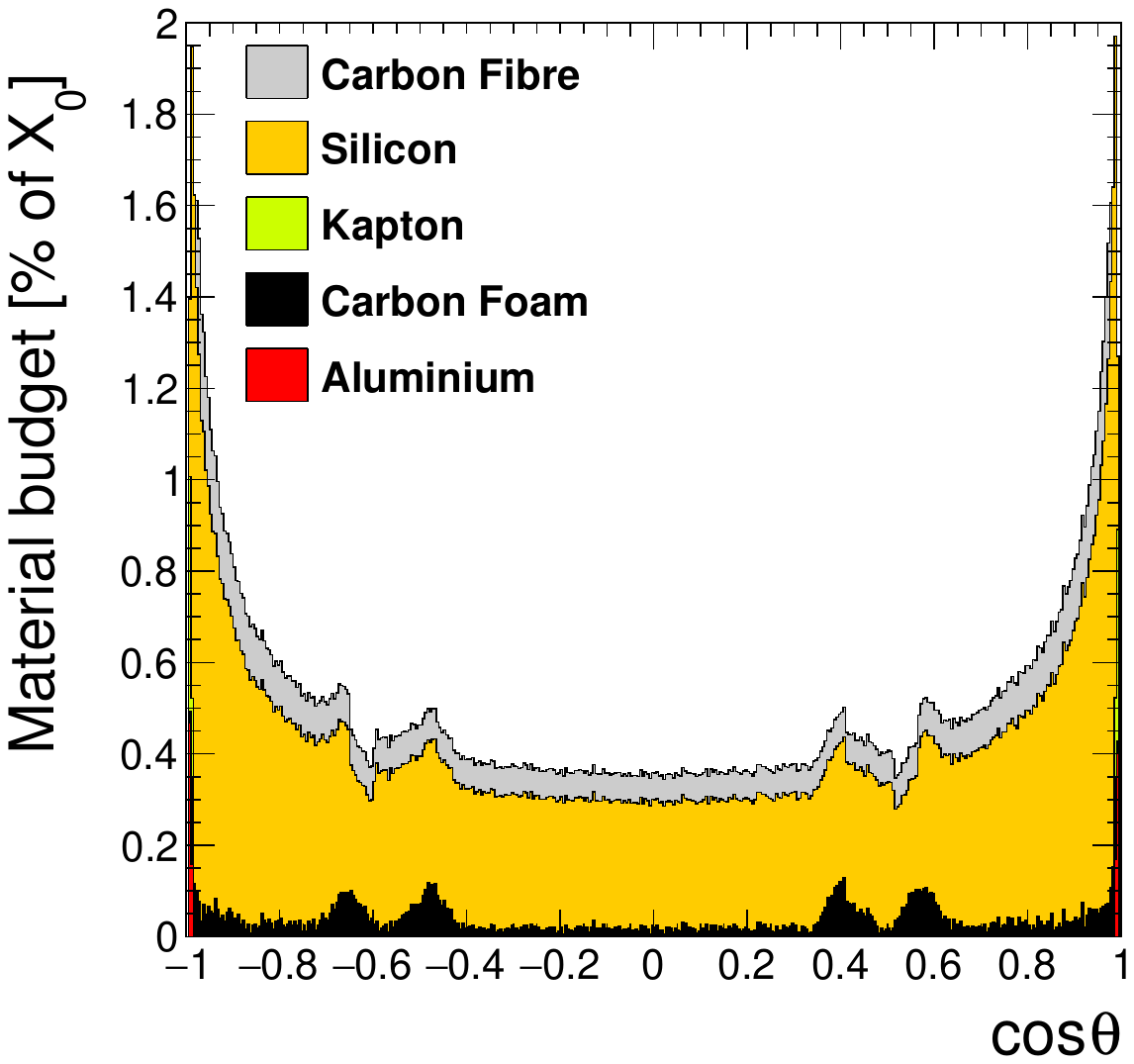}
    \caption{Material budget of the four-layer ultra-light inner VXD concept.}
    \label{fig:ultra-light_material_budget}
\end{wrapfigure}

Work has started to adopt such a design for FCC-ee VXDs \cite{Ilg2024}. \autoref{fig:ultra-light} shows the currently investigated layout featuring four curved inner vertex layers with a forward coverage of all layers up to $\left| \cos\theta \right| \approx 0.99$. Such a design can decrease the material budget of the inner VXD down to $\SI{0.075}{\percent}\ X_0$ per layer, which is more than a factor of three reduction. The material budget of the complete ultra-light inner VXD concept is displayed in \autoref{fig:ultra-light_material_budget}. It can be seen that cabling is limited to $\left| \cos\theta \right|> 0.985 $ and only light-weight carbon foam support structures (longerons and rings) and an outer carbon fibre enclosure are used.

A further improvement of vertexing capabilities is possible by tackling the beam pipe. The thickness and material budget of the current beam pipe design is already small, and further improvements are uncertain. One approach is to evaluate whether putting a first VXD layer before the beam pipe is possible, following the developments of LHCb \cite{lhcb_upgradeII} and ALICE 3 \cite{alice3eoi}. The challenges to overcome are the cooling the wake field-induced heat in the VXD and the significant hit rate expected without the beam pipe in front of the VXD. The impact on the beam dynamics needs to be studied as well.

The last front of VXD development is the timing capability. A time resolution of $\mathcal{O}(\SI{20}{ns})$ would maximise the track reconstruction efficiency as any hit could be assigned to the correct bunch crossing. One layer with a resolution of $\mathcal{O}(10\text{'s of } \SI{}{ps})$ could even serve as an inner time reference for time-of-flight measurements to enhance the particle-identification capabilities at FCC-ee further. 

The goal for the following years must be to assess the impact such VXD developments can have on physics performance and, thus, how they would extend the FCC-ee physics programme.

\begin{table}[htbp]
    \caption{Requirements of advanced VXD.}
    \label{tab:advanced}
    \centering
    \begin{tabular}{@{}p{0.355\textwidth}|p{0.64\textwidth}@{}}
        \textbf{Advanced challenges} & \textbf{Requirement} \\ \hline
        $\approx 2$ reduction of $\sigma_{d_0}$ & Smaller spatial resolution and $r_\text{min}$, lighter vertex and beam pipe  \\ \hline
        Bunch tagging/inner time reference & $\mathcal{O} (\SI{20}{ns})$ time resolution/$\mathcal{O}(\text{10's of ps})$
    \end{tabular}
\end{table}

\section{MAPS towards FCC-ee vertex detectors}

The various requirements of FCC-ee VXD listed above are stringent and often contradicting, with the material budget being the antagonist for all other requirements. Two examples of MAPS are discussed in the following that align with the direction of requirements for FCC-ee. The ARCADIA sensor \cite{Neubser2023} in the LFoundry \SI{110}{nm} CMOS process is currently used in the baseline IDEA VXD design. MOSAIX \cite{its3_tdr} is developed in the \SI{65}{nm} TPSCo CMOS process and will be used for ALICE ITS3. Both devices feature a power consumption compatible with FCC-ee, while the sensor spatial resolutions are only $\approx \SI{5}{\micro\meter}$. MOSAIX (ARCADIA) is designed to be capable of dealing with hit rates of 10 (100) \SI{}{\mega\hertz\per\centi\meter\squared}, which still is below the currently assumed requirement for FCC-ee. Stitching is used to build wafer-scale (side-abuttable) sensors and a \SI{2}{\micro\second} integration time ($\mathcal{O}(\SI{}{ns})$ time resolution) is achieved in MOSAIX (ARCADIA). In both processes, studies are ongoing to improve the time resolution to hundreds or tens of picoseconds, at the cost of a significant increase in power consumption. It can be concluded that no MAPS prototype that can fulfil all requirements exists yet. Examples of MAPS developments towards FCC-ee VXDs are CE-65 and H2M, both produced in the \SI{65}{nm} TPSCo process and presented at the Pixel2024 workshop \cite{Lorenzetti2025,daza2025h2mmonolithicactivepixel}.

\section*{Conclusions}

The performance and environmental challenges that FCC-ee VXD detectors face are stringent and numerous. R\&D is needed on all fronts, from mechanics and MDI integration to MAPS development and data acquisition. In addition to the development of ultra-light and fast detectors discussed above, novel technologies such as smart on-detector processing, power over fibre, and wireless readout must also be investigated. The final goal must be to minimise the systematic uncertainties in the measurements to benefit fully from the large datasets at FCC-ee.

\bibliographystyle{JHEP}
\bibliography{biblio.bib}

\end{document}